# Collision statistics of clusters: From Poisson model to Poisson mixtures

**Sascha Vongehr, Shaochun Tang (唐少春), and Xiangkang Meng (孟祥康)**

National Laboratory of Solid State Microstructures, Department of Materials Science and Engineering, Nanjing University, Nanjing 210093, P.R. China

Abstract: Clusters traverse a gas and collide with gas particles. The gas particles are adsorbed and the clusters become hosts. If the clusters are size selected, the number of guests will be Poisson distributed. We review this by showcasing four laboratory procedures that all rely on the validity of the Poisson model. The effects of a statistical distribution of the clusters' sizes in a beam of clusters are discussed. We derive the average collision rates. Additionally, we present Poisson mixture models that involve also standard deviations. We derive the collision statistics for common size distributions of hosts and also for some generalizations thereof. The models can be applied to large noble gas clusters traversing doping gas. While outlining how to fit a generalized Poisson to the statistics, we still find even these Poisson models to be often insufficient.

## 1. Introduction

When clusters traverse gas, they collide with the gas particles. For a cluster of a given size, the random collisions are described by a Poisson process, but the size distribution of clusters inside a beam is usually broad. While fractionation leads often to exponential distributions, random growth processes like phase change aggregations give mostly rise to log-normal ones, be it in biology, economics or cluster physics [1,2,3]. Usually, in beams of clusters, the clusters' number of atoms $N$ has a standard deviation almost as large as the average size. This large initial uncertainty complicates the statistics of the collision experiment. The absolute differences of more than 100% between probabilities in simple and improved models (figures 1 to 3) are mostly not practically relevant. We draw from laboratory experience and only want to point out practically important corrections. Therefore, we review the simple Poisson model (section 2) by showcasing four procedures that rely on the validity of that model: Single molecule isolation spectroscopy (HENDI [4,5]), the dependence of mass spectrometric peaks on changing doping gas density, and so on. With the desire to be practically relevant, we first provide formulas for the influence of the host cluster distribution on the capture statistics. Firstly, we derive analytically exact expressions for the effects on average capture rates (section 3). These sections also introduce a necessarily strict notation. Statistical quantities like average and variance all exist for the probability distribution of the hosts and also for that of the guests. Then we introduce Poisson mixtures and derive exact expressions for the standard deviations[a] (section 4) from the Poisson mixtures' general properties.

Reporting a measurement via its average and standard deviation is the accepted scientific standard. Why do we deal with the full complexity of Poisson mixtures (section 5 to 7)? Our initial motivation was to model exotic collision and pick-up cross sections [6] due to electrical polarizability, sticking coefficients, etc. However, the mixture models had already unexpected results when merely still considering simple cross sections. Sometimes, the shape of the actual distribution can mislead the researcher's interpretations about physical processes (especially section 5 and 7). Sometimes, the average and deviation shift only a little, but application of the Poisson model leads nevertheless to significant errors (section 6).

---

[a] A downwards correction of an average without correcting the deviation necessarily increases the height of the maximum. Actually, the maximum is always lower (see black peaks, figures 1 to 3).

The treatment is only valid for large hosts picking up guests that do not provide too much energy (e.g. via condensation to a strongly bound guest cluster). Especially when dealing with small host clusters, it is necessary to account for the evaporation resulting from dissipation of impact and binding energy and more subtle effects on the final host cluster size distribution resulting from preferential scattering of small clusters out of the beam. Such can only be done numerically, for example with a Monte Carlo modeling method [7,8], but it is difficult to infer conclusions valid under different conditions without running simulations every time again, especially if a detection method (laser induced fluorescence (LIF), ionization, etc) is already nested inside the simulation. Evaporation renders collisions history dependent, i.e. the next collision depends on how many there have been already. This violates the main assumption of a Poisson process. Hence, the widespread use of so called effective cross sections while keeping the Poisson model cannot be a mathematically sound approach. Moreover, we show that the deviation needs to be decoupled from the average. There are many statistical distributions that do so. It is our strategy to select the most promising ones by sticking to analytical models with well understood parameters for as long as possible. One eventually must give up analyticity and maybe also enter effective cross sections in order to account for evaporation, detector parameters etc. Nevertheless, the strategy demands to first correctly account for the influence of the hosts' size distribution. It seems to be the only effect that can be dealt with analytically. Moreover, this clearly distinguishes the specific influence of the hosts' size distribution on the shape of guest distributions. Beam depletion by deflection and evaporation are very important for small hosts, but with large hosts, the effects of distribution mixing are more important. In other words, we distill out the effects of the easily neglected large size "tail" of the host distributions. They turn out surprisingly important.

All probability distributions and probability density functions (PDF) that we used for calculations are listed in the appendix in order to facilitate comparison.

## 2. The Poisson model and its practical use

For clarity, we will present everything while having in mind large noble gas clusters that capture collision partners efficiently and are not deflected much. In Helium nano-droplets for example, guest particles will condense very fast, and after $k$ collisions, the host cluster contains a guest cluster of size $k$. Generally, when a host travels through a length of gas, the probability of colliding with $k$ particles is $P_k$. The probability *not* to collide is $P_0 = e^{-\bar{k}}$. It depends[b] on $\bar{k} := D\sigma$, where $\sigma$ is the cross section and $D$ the doping strength. $D := \tilde{n}FL$ combines the gas related properties: particle number density $\tilde{n}$, path length $L$, and $F$, which takes into account the velocity distribution[c] of the gas [9]. If collisions are independent of each other and the clusters are massive enough to collide with several particles without being deflected, one can use $P_0$ to straightforwardly deduce the Poisson distribution[d] $P_k = P_0 \bar{k}^k / k!$. This distribution is equi-dispersed, i.e. the variance equals the mean:

$$\text{var} := \sum_k (k - \bar{k})^2 P_k = \bar{k} \tag{1}$$

The Poisson distribution has therefore only one degree of freedom and is rather rigid.

---

[b] ":=" defines a new symbol on its left hand side, while "=:" defines a new one on the right side.

[c] $\sqrt{\pi} \, F_{(x)} = x^{-1} e^{-x^2} + (2 + x^{-2}) \int_0^x e^{-t^2} dt$ ; $x = v/\hat{u}$. The host cluster's speed is $v$ and the most probable speed $\hat{u}$ of the scattering gas particle is given via $m\hat{u}^2 = 2k_B T$.

[d] The bar over $k$ does not extend over the power $k$. A bar always indicates the averaging over all $k$.

There are four methods of practical importance in the laboratory that all rely on the Poisson model being valid. We will revisit them often later on in order to show how the different host size distributions may have an impact on each of these methods:

*1: Monomer isolation:* Noble gas host cluster beams are used for ultra cold isolation spectroscopy. Typically, one investigates isolated molecules via their LIF signal. The doping is chosen to be so weak that the average number of picked up molecules is below unity. With $\bar{k} \cong 0.3$ for example, only about 25% of hosts pick up anything at all ($k > 0$). For every host that picks up one guest, only $P_2/P_1 = \bar{k}/2 \cong 0.15$ collide with two. The LIF is then treated as if originating from isolated molecules only. Of all doped hosts, only $(1 - P_0 - P_1)/(1 - P_0) \cong 14\%$ carry clusters instead of the desired monomers.

*2: Weak doping dependence* (WDD): When few collisions occur, the probability can be expressed as the WDD limit $\lim_{\bar{k} \to 0} P_k = \bar{k}^k (1 - \bar{k})/k!$. A signal due to guests of size $k$ is proportional to $P_k$ and therefore a function of the respective average to the $k^{\text{th}}$ power at the origin ($\bar{k} = 0$). A signal's linear rise (i.e. one to the 1$^{\text{st}}$ power) accompanying a slight change in the doping gas pressure is often used to argue that the signal is due to monomers $k = 1$ only. A quadratic rise may identify a signal, such as a certain LIF frequency, as originating from dimers (a cluster with two monomers). One may also observe small guest clusters and infer the involved cross sections by employing $\lim_{\bar{k} \to 0} P_1 = \bar{k} - \bar{k}^2$ and $\lim_{\bar{k} \to 0} P_2 = (\bar{k}^2 - \bar{k}^3)/2$.

*3: Maximum when changing doping* (MCD): The derivative $\frac{dP_k}{d\bar{k}} = \frac{P_k}{\text{var}}(k - \bar{k})$ shows that $P_k$ is maximal when the average pick-up is $\bar{k} = k$. This allows deriving total cross sections [10] after finding the doping strength that results in a maximum signal due to hosts of size $k$.

*4: Fixed doping ratios* (FDR): *Monomer isolation* tries to avoid the total of all multi-mer signals because one may not be able to relate signal type to guest cluster size. Now think of signals with known origin (guests of size $k$) and of larger collision rates. Consider the mass abundance spectrum of guest clusters. Given one peak in the mass spectrum, the FDR ratio $P_k = (\bar{k}/k) P_{k-1}$ predicts all other peaks. The highest peak (the maximum around $P_{k-1/2} = P_{k+1/2}$) is found at $k \cong \bar{k} - 1/2$.

### 3. Average collision rates

The host clusters' sizes $N$ are statistically distributed. Since one observes a statistical ensemble of hosts in the beam, the expectation value of any observable $\Psi$ is[e] $\langle \Psi \rangle = \int_0^\infty \Psi \, \text{PDF} \, dN$. A linear exponential (EXP) with $\text{PDF}^{-1} = \langle N \rangle e^{N/\langle N \rangle}$ has been observed for large clusters gained from so-called supercritical expansions, e.g.: for Helium droplets [11] $He_N$. The EXP is equi-dispersed, i.e., the standard deviation equals the mean $\Delta_N = \langle N \rangle$. It therefore has only one degree of freedom. Whenever possible, we derived all desirable equations for more general distributions dependent on the dispersion ratio, which is defined as:

$$d_X := \langle X \rangle^2 / \Delta_X^2 \qquad (2)$$

For instance, results for the EXP follow at once by setting $d_N = 1$ in formulas valid for the gamma ($\Gamma$) distribution. After sub-critical beam expansions, the condensed clusters have log-normal (LN) size distributions whose two degrees of freedom are given via the mean $\langle n \rangle$ and standard deviation $\Delta_n$ of the logarithm $n := \ln N$. The LN is simply the well known normal distribution in $n$-space.

---

[e] Notation: An average over all $N$ uses angled brackets, e.g. $\langle P_k \rangle$. Dependence on discrete variables is written without brackets, as for $P_k$. Dependence on a continuous variable uses round brackets, as in $\sigma_{(N)}$. $N$ is continuous because the present work is about very large clusters. There is no convenient discretization of the host size distributions considered here. Variables may serve as identifying (rather than counting) index, e.g. the standard deviation $\Delta_N := \sqrt{\langle N^2 \rangle - \langle N \rangle^2}$.

The following formula is very useful, especially when closed expressions are impossible and one needs to expand an observable $\Psi$ in a power series:

$$\langle N^a \rangle = \langle N \rangle^a \begin{cases} d_N^{-a} \Gamma_{(d_N+a)}/\Gamma_{(d_N)} & \text{; for } \Gamma\text{- distr.} \\ \exp\left[a(a-1)\Delta_n^2/2\right] & \text{; for LN-distr.} \end{cases} \quad (3)$$

An important example is a cluster's geometrical cross section $\sigma = \pi r_S^2 N^{2/3}$ with the Wigner-Seitz radius $r_S$ (e.g.: $r_{S-\text{Helium}} = 2.221\text{Å}$) [12]. Since $\bar{k} \propto \sigma = \sigma_{(N)}$, a shift of $\langle \sigma \rangle$ will shift the expected average number of collisions or guests $\kappa := \langle \bar{k} \rangle = D \langle \sigma \rangle$. Thus, FDR and the MCD both shift. In case of the EXP-distribution, the observed average is only 90.3% as large as naively expected, because of $\langle \sigma \rangle / \left(\pi r_S^2 \langle N \rangle^{2/3}\right) = \Gamma_{(5/3)} \approx 0.903$. This large correction holds for any exponential size distribution, regardless of its width $\Delta_N$. Cluster beams can usually be described via $\Delta_N \cong \text{few} \langle N \rangle$; the LN-distributed ones, too. For Helium clusters from continuous nozzles it is experimentally established that they obey [13,14]

$$\Delta_n = 0.55 \pm 0.03 \quad \Rightarrow \quad d_N \cong 2.83 \quad (4)$$

This leads to a correction of $e^{-(\Delta_n/3)^2} = 96.7\%$. This is not as impressive as the 10% error for the EXP. However, it will be shown later how even this small deviation can lead to a discrepancy of almost 30%. This comes about because the new average is actually that of a differently shaped distribution. The small shift here does sadly not imply that the Poisson distribution can be trusted for LN-distributed hosts. Also, the corrections do not depend on the doping. They are just as valid in the limit of vanishing gas density and should be taken into account even when observing the WDD.

### 4. Poisson mixtures

The just discussed average number of collisions or guests $\kappa = \langle \bar{k} \rangle$ is only the first important degree of freedom of the guest clusters' distribution $\langle P_k \rangle$, which is called a Poisson mixture [15]. Poisson mixture models are well known in actuarial science to model total insurance claim distributions. The Poisson mixture is said to be "mixed" by the hosts' size distribution. The size distribution is called "mixing distribution". Using $p \in \{0,1\}$ for the normalization and the average of a distribution respectively, un-mixed and mixed expressions are very similar, i.e., $\sum_k k^p P_k = \bar{k}^p$ becomes $\sum_k k^p \langle P_k \rangle = \kappa^p$. The mixture's variance is $\mathrm{VAR} := \sum_k (k-\kappa)^2 \langle P_k \rangle$. A Poisson mixture is always over-dispersed because the Poisson distribution's $\mathrm{var} = \bar{k}$ leads to the mixture's variance[f] being

$$\mathrm{VAR} = \kappa + \Delta_{\bar{k}}^2 = \kappa\left(1 + \kappa/d_{\bar{k}}\right) \tag{5}$$

This allows us to present correction formulas for deviations even when lacking a closed expression for the mixed distributions. The LN is convenient: Given an LN host size distribution, any proportional variable $M \propto N^a$ is also log-normal with $\Delta_{\ln M} = a \Delta_n$ and $\langle \ln M \rangle = a \langle n \rangle + \ln(M/N^a)$. Defining $\bar{k} =: B N^a$ leads thus straight to $\Delta_{\bar{k}}$ and $d_{\bar{k}}$. For the EXP, the distribution of $\bar{k}$ and thus $d_{\bar{k}}$ can be derived with the cumulative distribution function of the EXP. The results are

$$d_{\bar{k}}^{-1} = -1 + \begin{cases} 4^a \Gamma_{(a+1/2)} / (a!\sqrt{\pi}) & \text{; for EXP-distr.} \\ \exp\left[\Delta^2_{\ln \bar{k}}\right] & \text{; for LN-distr.} \end{cases} \quad (6)$$

---

[f] Changing the sequence of sum and integration ($\sum \langle ... \rangle = \langle \sum ... \rangle$) and several times applying $\mathrm{var}(x) = \mathrm{mean}(x^2) - (\mathrm{mean}(x))^2$ shows $\mathrm{VAR} = \langle \mathrm{var} \rangle + \Delta_{\bar{k}}^2$ holds true for any mixture model.

Even when lacking closed expressions for the mixtures, one can still derive some further general statements by looking at the power expansions. Via expanding

$k!\langle P_k \rangle = \langle \bar{k}^k e^{-\bar{k}} \rangle = \sum_{i=0}^{\infty} (-1)^i \langle \bar{k}^{k+i} \rangle / i!$ one may derive a general expression for Poisson mixtures (10) and investigate the leading terms.

The WDD has the same leading power after mixing (i.e. after averaging over the hosts' sizes), but to infer total cross sections from signals of small guest clusters at weak doping is not so easy anymore: The proportionality factors for $\lim_{\kappa \to 0}\langle P_{k\geq 2} \rangle$ and for the terms in $\lim_{\kappa \to 0}\langle P_1 \rangle$ of higher than first order in $\kappa$ depend on the size distribution of the hosts:

$$\lim_{\kappa \to 0}\langle P_k \rangle = \frac{\kappa^k}{k!} \begin{cases} \left[\Gamma_{(1+ak)} - \kappa \Gamma_{(1+a(k+1))}/\Gamma_{(1+a)} + ...\right] / \Gamma^k_{(1+a)} & \text{; for EXP-distr.} \\ \left[1 - \kappa e^{k\Delta^2_{\ln \bar{k}}} + \kappa^2 e^{(2k+1)\Delta^2_{\ln \bar{k}}}/2 - ...\right] * e^{\frac{k}{2}(k-1)\Delta^2_{\ln \bar{k}}} & \text{; for LN-distr.} \end{cases} \quad (7)$$

For the LN, this can be more intuitively expressed with $\varepsilon := (1+1/d_{\bar{k}})$, because one can then write $\lim_{\kappa \to 0}\langle P_1 \rangle = \kappa\left[1 - \varepsilon\kappa + \varepsilon^3\kappa^2/2 - ...\right]$ and $\lim_{\kappa \to 0}\langle P_2 \rangle = \frac{\kappa^2}{2}\varepsilon\left[1 - \varepsilon^2\kappa + \varepsilon^3\kappa^2/2 - ...\right]$.

The average of collisions $\kappa$ may also be varied by adjusting the average host size $\langle N \rangle$ instead of the doping gas pressure, but in practice this alters the host clusters' beam expansion conditions, which changes the overall beam flux. Therefore, we let $\langle N \rangle$ untouched and that is how the derivative $d\langle P_k \rangle/d\kappa$ should be understood here. Using $d/d\kappa = \langle N^a \rangle^{-1} d/dB$, one may drag the derivative into the integral $\int P_k \text{PDF} dN$, where it can only act on $P_k$, because the PDF is independent of doping. It follows that the MCD is at

$$k \langle P_k \rangle = (k+1) \langle P_{k+1} \rangle \tag{8}$$

This concludes the general part. Even simplifying linear approximations (e.g.: $\bar{k} = BN^a$ with $a = 1$) are generally not helpful. Obtaining a closed, let alone tractable expression for $\langle P_k \rangle$ is possible only for certain mixing distributions [11,16].

### 5. Instructive example: linear cross section with exponential (gamma) mixing

A linear approximation $\bar{k} = BN^a$ with $a = 1$ may be physically due to the cross section being (effectively) proportional to $N$. This leads to $\kappa = B\langle N \rangle$. Since it is easily done and needed later, the results are here given for the hosts' sizes being $\Gamma$-distributed; the results for the EXP follow by setting $d_N = 1$. The Poisson mixture $\langle P_k \rangle$ is the probability function of the Negative Binomial (NB) or "Pascal" distribution (11). For $d_N = 1$ it is called "geometric" distribution. The variance is $\text{VAR} = \kappa(\kappa + d_N)/d_N$. Apart from the difference between var and VAR, the derivative retains the form it had before the mixing: $\dfrac{d\langle P_k \rangle}{d\kappa} = \dfrac{\langle P_k \rangle}{\text{VAR}}(k - \kappa)$. Therefore, relative to $\kappa$ instead of $\bar{k}$, the MCD survives the mixing. The WDD becomes:

$$\lim_{\kappa \to 0} \langle P_k \rangle = \frac{\kappa^k}{k!} \frac{\Gamma_{(d_N + k)}}{(d_N + \kappa)^k \Gamma_{(d_N)}^k} (1 - \kappa) \tag{9}$$

For clarity and practical significance, the rest of this section assumes an EXP distribution. The limit $\lim_{\kappa \to 0} \langle P_k \rangle = \kappa^k$ confirms that the WDD of $\langle P_1 \rangle$ is the same after the mixing (i.e. after averaging over host sizes) as it was before, at least to first order in $\kappa$. However, all $\langle P_k \rangle$ with $k>1$ are modified: $\lim_{\kappa \to 0}\langle P_k \rangle = \kappa^k (1-\kappa(k+1))$. *Monomer isolation* is compromised because for every host that collides only once, $(1-\langle P_0 \rangle - \langle P_1 \rangle)/\langle P_1 \rangle = \kappa$ pick up multiple guests. With $\kappa = \overline{k} \cong 0.3$, now 23% of the doped hosts carry clusters instead of the desired monomers. The FDR becomes $\langle P_k \rangle = (1-\langle P_0 \rangle)\langle P_{k-1} \rangle$. This implies that a mass spectrum of the guest clusters looks now always as if the signals are systematically suppressed. Peaks proportional to $\langle P_k \rangle$ monotonically decrease with $k$.

In order to illustrate the effects with an example, assume that the doping strength (gas density) and the average size of the hosts lead together to on average $\kappa = \overline{k} = 10.5$ guests. To compare the traditional with the improved models, we calculated guest abundance spectra (figure 1). The Poisson model (grey) $P_k = \frac{10.5^k}{k!} e^{-10.5}$ predicts the maximum at $k=10$. Taking the exponential host cluster size distribution into account, the spectrum (black) is due to a geometric distribution $\langle P_k \rangle = \left(\frac{10.5}{11.5}\right)^k \frac{1}{11.5}$. It looks as if it is obtained from a violent fractionation of very large guest clusters. There is no maximum anymore. Instead, a strong right tail with unexpectedly large guests appears. The effect is here especially obvious because of the simplified cross section and the EXP size distribution. However, this effect of the host's size distribution mixing is always present to some extend. This generally leads to overestimation of the severity of fractionating detection processes or host depletion due to binding energy release when guests meet in/on the host.

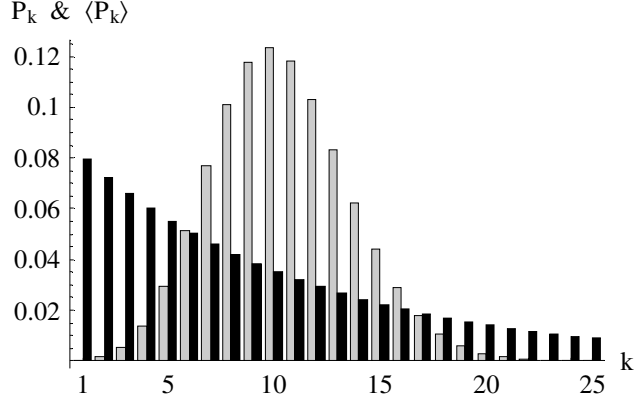

Figure 1: Guest abundance spectra (= collision statistics) due to the Poisson model (gray) and due to a Poisson mixture (black). A shape typical of growth processes turned into one typical for fractionations. This illustrative example assumes a simplified cross section.

## 6. Log-normal host size distribution

To mix a Poisson with an LN is difficult. As long as $\Delta_n \leq 0.6$ (compare equation (4)), the Inverse Gaussian (IG) distribution [17,18] traces the LN extremely well. Considering the experimental accuracy, sub-critical cluster size distributions could have been modeled with the IG all along. The IG may also be as well supported as the LN when considering the underlying physical processes. We would like to draw attention to the fact that the IG may be very useful for cluster physics. Its Poisson mixture is a closed expression.

We want to fit the distribution of $\bar{k} = D\sigma$. The cross section $\sigma$ is proportional to $N^a$, so their distributions are also log-normal with $\Delta_{\ln \bar{k}} = \Delta_{\ln \sigma} = a\Delta_n$. Since $a \cong 2/3$, it follows from equation (4) that $\Delta_{\ln \bar{k}} \cong 0.37$ is small. Hence, one may with confidence express the ensemble of host clusters via the IG's $\text{PDF}_{(d_{\bar{k}}, \kappa)}$ while using equation (6), namely $\kappa = \exp\left[\frac{1}{2}\Delta_{\ln \bar{k}}^2 + \langle \ln \bar{k} \rangle\right]$, $\Delta_{\ln \bar{k}} = a\Delta_n$ and $\langle \ln \bar{k} \rangle = a\langle n \rangle + \ln B$. The collision probability can be integrated. The probability of no collision is $\langle P_0 \rangle = \exp\left[d_{\bar{k}} - \sqrt{d_{\bar{k}}\Omega}\right]$ with $\Omega := d_{\bar{k}} + 2\kappa$. The general formula

(12) is now the probability function of the IG Poisson mixture distribution (IGP). In the weak doping limit, the following expression is valid for general $k$:

$$\lim_{\kappa \to 0} \langle P_k \rangle = \left(\frac{\kappa}{\Omega}\right)^k d_{\bar{k}}^{k-\frac{1}{2}} \sqrt{\frac{2}{\pi}} \frac{e^{d_{\bar{k}}}}{k!} \left( (d_{\bar{k}} + \kappa) K_{\frac{1}{2}-k}(d_{\bar{k}}) - \kappa d_{\bar{k}} K_{\frac{3}{2}-k}(d_{\bar{k}}) \right).$$

E.g. for the probability of only one collision one may write $\lim_{\kappa \to 0} \langle P_1 \rangle = \kappa \left[ 1 - \varepsilon \kappa + \frac{1}{2}\left[ \varepsilon^3 - (1/d_{\bar{k}})^3 \right] \kappa^2 - ... \right]$, which is the same as the result for the LN up to $2^{nd}$ order in $\kappa$ (compare equation (7) and thereafter). Thus, the main difference to the WDD between Poisson model and LN mixed one is captured correctly by the IGP. Already $\lim_{\kappa \to 0} \langle P_2 \rangle = \kappa^2 \varepsilon$ is twice the LN's result. This could perhaps be measured to justify preferring the IG over the LN distribution.

Only in the limit $d_{\bar{k}} = \infty$ is the MCD still at $\kappa = k$. Otherwise it is at

$$\frac{d\langle P_k \rangle}{d\kappa} = \frac{\langle P_k \rangle}{\kappa \Omega} \left( (d_{\bar{k}} k + \kappa) - \frac{\kappa \sqrt{d_{\bar{k}} \Omega} K_{\frac{3}{2}-k}(\sqrt{d_{\bar{k}} \Omega})}{K_{\frac{1}{2}-k}(\sqrt{d_{\bar{k}} \Omega})} \right).$$

The probability of a single collision is therefore $\frac{d\langle P_1 \rangle}{d\kappa} = \frac{\langle P_1 \rangle}{\kappa \Omega}\left( d_{\bar{k}} + \kappa - \kappa \sqrt{d_{\bar{k}} \Omega} \right)$. For example, at $d_{\bar{k}} = 2$, the maximum of $\langle P_{1,2,3} \rangle$ is shifted by 6.7, 10.8 and 13.7% away from $\kappa = 1, 2, 3$ respectively, requiring a *higher* average doping to reach the maximum value. The shift upwards from $\kappa = k$ worsens with $k$ approaching a limit of 27.9%.

Guest cluster abundance spectra (figure 2) for hosts with the geometrical cross section of liquid drops, i.e. $a = 2/3$, and a doping strength leading to on average 10 guests per host were calculated: Uncorrected, the average value is assumed too high ($\bar{k} \cong 10.34$) and the Poisson model (white) predicts a maximum at $k = 10$. The Poisson model with a corrected average (grey) of $\kappa = \langle \bar{k} \rangle$, here $\kappa = 10$, predicts the peaks at $k = 9$ and $k = 10$ to be equal. In the actual spectrum due to the IGP (black), the peak at $k = 8$ is the highest. Also, a strong right tail appears again.

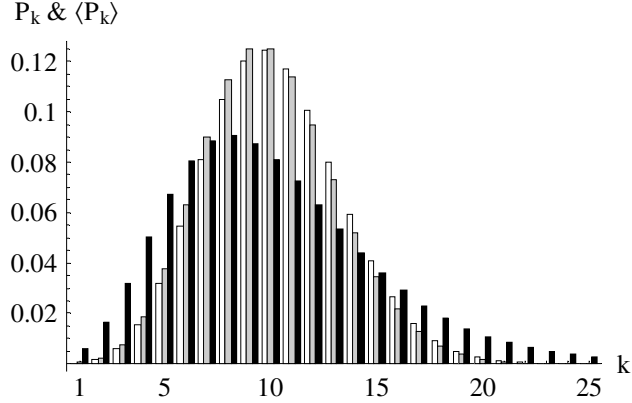

Figure 2: Guest abundance spectra due to the uncorrected Poisson model (white), the model with corrected average (gray), and the Poisson mixture (black) that takes the LN host size distribution into account. Fitting the latter with a Poisson curve would result in large errors.

*Monomer isolation* survives: With $\bar{k} = 0.3 \Leftrightarrow \kappa = 0.29$, $\langle P_2 \rangle / \langle P_1 \rangle \cong 0.159$ is close to $P_2/P_1 = 0.15$. Now 15% of the doped hosts carry clusters instead of monomers. The corrections are less severe than with the EXP distribution, because the LN is under-dispersed (equation (4) leads to $d_{\bar{k}} \approx 6.95$). The smallness of the power $a$ cannot[g] be blamed. Hence, one may be tempted to interpret the smallness of corrections as reassuring. The Poisson model seems only unable to deal with EXP distributed host sizes, but this is deceptive. What is routinely done [19] is to fit the guests' abundance spectrum with a Poisson distribution in order to infer the size of the hosts. For the calculated case (figure 2), such fitting underestimates the average pick-up by almost 20%, because a Poisson's average and maximum are close and the maximum peak went from 10 down to $k = 8$ during the mixing with the hosts' size distribution. Calculating the average number $\langle N \rangle$ of atoms in the hosts introduces the power $a = 2/3$. This in turn worsens the matter to only $(80\%)^{3/2} \cong 72\%$ of the actual average. It is true that taking the LN into account reduces the average number of collisions $\kappa$ to about 96.7%, but one should not conclude that utilization of the Poisson model will only lead to errors of about 4%.

---

[g] One may put $a = 1$ again: In the IGP-spectrum with $\kappa = 10$ (Fig. 2), $\langle P_6 \rangle$ would be the highest, so the linear approximation, compared to $a = 2/3$, only doubles the shift from the uncorrected $k = 10$. With $a = 1$ and thus $\kappa = \bar{k}$, *monomer isolation* at $\kappa = 0.3$ would have 17.6% of doped hosts carry clusters.

## 7. Exponential host size distribution

If the power $a < 1$, the distribution of $\bar{k}$ will have a cusp around the maximum even though $N$ is EXP distributed; shifts will be less severe than those of section 5. In order to calculate the Poisson mixture, we need to fit the PDF of $\bar{k}$ again. We looked at many possibilities: The Power Inverse Gaussian [20,21] (PIG) is attractive because the LN belongs to it as a special limiting case. The PIG can fit the distribution of $\bar{k}$ for both types of size distributions (!), but it has no convenient Poisson mixture. The Generalized Inverse Gaussian distribution (GIG) includes many others as special cases ($\Gamma$, Hyperbolic, Reciprocal Inverse Gaussian (RIG), …) and allows tractable Poisson mixture [22]. Of the whole GIG family, $\Gamma$ fits the PDF of $\bar{k}$ best, but, when sharing the same $d_{\bar{k}}$ and $\kappa$, the $\Gamma$ is more peaked[h] than the distribution of $\bar{k}$. After fitting, one can use the results for the $\Gamma$ with linear approximation from above to calculate an NB distribution. However, treating the LN taught us that a more peaked function can strongly underestimate the properties of interest here. Being presented with these difficulties, we abandoned generality and only calculated for researchers in cluster science most important conditions.

---

[h] A different fit (e.g. a least squares fit) does not make the GIG fit well. Moreover, functional fits match average and dispersion and thereby preserve those important quantities.

The exact Poisson mixture (equation (13)) for the hosts having an EXP size distribution and $a = 2/3$ leads to the following results: The MCD is shifted *downwards*, i.e. it is reached already at lower than expected gas pressures. The maxima of $\langle P_{1,2,3,...,10} \rangle$ are shifted by 3.0, 4.6, 5.6, 6.2, 6.7, 7.0, 7.3, 7.6, 7.8 and 7.9% away from $\kappa = 1, 2, 3, ..., 10$ to a lower average doping. *Monomer isolation* is somewhat compromised: With $\bar{k} = 0.3 \Leftrightarrow \kappa = 0.27$ and $\langle P_2 \rangle / \langle P_1 \rangle \cong 0.178$, now 17% of the doped hosts carry clusters instead of isolated monomers. Let us again compare guest cluster abundance spectra. We assume that the experimental conditions lead to on average $\kappa = 9.5$ guests (figure 3): Uncorrected, the average is assumed too high ($\bar{k} \cong 10.52$) and the maximum is predicted to be at $k = 10$. A corrected Poisson model still predicts the peak at $k = 9$ to be the highest. Taking the exponential distribution fully into account, the peak at $k = 4$ wins, a strong right tail with unexpectedly large clusters appears, and the overall shape suggests fractionation processes. The Poisson model fails because it fixes the maximum to be close to the average.

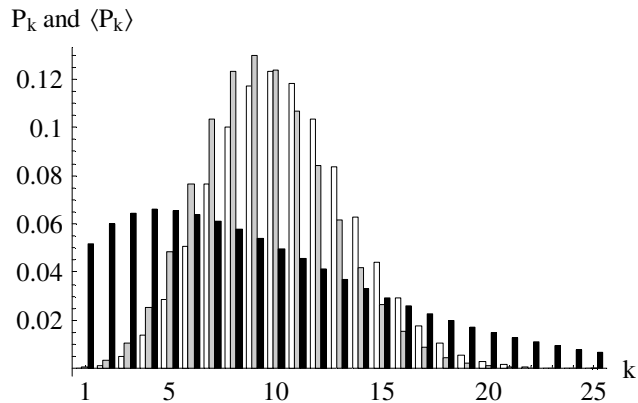

Figure 3: Guest abundance spectra due to the Poisson model (white), the model with corrected average (gray), and the Poisson mixture (black) that takes the EXP host size distribution into account.

## 8. Generalized Poisson distribution

To simply facilitate a fitting of the shape of the collision statistics, one should bear in mind that any mixing will over-disperse an originally equi-dispersed distribution. In other words, the deviation needs to be decoupled from the mean to allow one more degree of freedom. A well-studied alternative to the standard is the generalized Poisson [23] distribution (GP). It has the mean $\bar{k} = u/(1-\lambda)$ and $\text{var} = u/(1-\lambda)^3$. It is over dispersed when $\lambda > 0$ and reduces to the Poisson at $\lambda = 0$. Could a functional fit with a GP replace the Poisson mixtures? To find out, we rewrite the GP as if it is already a mixture (14), i.e. we substitute $P_k \to \langle P_k \rangle_{GP}$ and put in the mixture's average instead ($\bar{k} \to \kappa$). We fix $\lambda$ in the new expression by requiring also the variance $\text{var} \to \text{VAR} = \kappa/(1-\lambda)^2$ to be correct, i.e. equal to the Poisson mixture's (5). This implies $(1-\lambda)^{-2} = 1 + \kappa/d_{\bar{k}}$. In order to present an example (figure 4) resembling those above, doping and average host size lead to on average 9.5 guests. The envelope of the guests' spectrum due to an EXP-Poisson mixture (black lines) is not well reproduced by the GP (dashed black line). Such fits may be sufficient with LN distributed host sizes (grey lines).

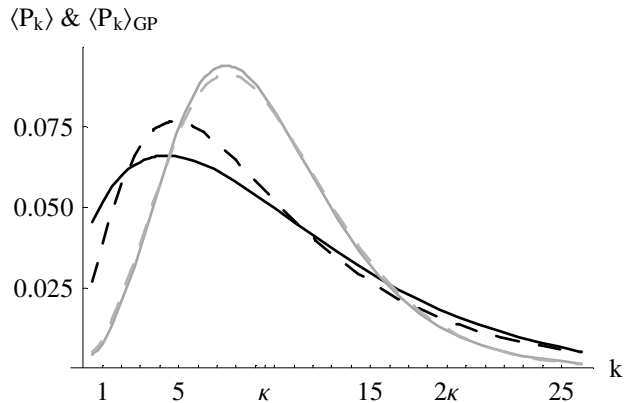

Figure 4: The envelopes of guest abundance spectra. The LN and EXP host size distributions are taken into account (grey and black lines respectively). The Poisson mixtures (solid lines) are each fitted with a generalized Poisson (dashed lines).

## 9. Conclusion

We calculated the collision statistics of size distributed host clusters as far as possible analytically and without use of effective cross sections. The resulting distributions allow a broader standard deviation, but there is a large pool of probability distributions that do so. In other words, the restriction to employ only well understood parameters enabled us to select specific probability distributions. This mathematical rigor surprisingly disfavors the well known generalized Poisson model. Host size distributions can be different from the most usual ones. Moreover, effective pick-up cross sections may depend on a power $a$ different from the geometrical $a = 2/3$ size dependence, for example when considering charged particles, guest desorption [6] or cluster surface layer corrections [24]. Therefore, we provided the Poisson mixture models by keeping expressions as general as possible. Nevertheless, with the desire to stay practically relevant, common laboratory methods were discussed and relatively simple correction formulas for estimation of means and standard deviations have also been provided.

It was shown that even for LN distributed cluster sizes, knowing average number of collisions and standard deviation can be insufficient. Taking the LN into account reduces the average number of collisions only to 96.7%. However, it is wrong to conclude that usage of the standard Poisson model will only lead to errors of maximally 4%. A simple example showed how the Poisson model can lead to an error of almost 30% instead. We have also shown that if the beam's host clusters have sizes that are EXP distributed, the Poisson model will be often insufficient. Under these conditions, monomer isolation will be compromised. An important result concerns the interpretation of the distributions shape: Even if the distribution of the host clusters does not change the expectation values much, the shape of a mass abundance spectrum can suggest a fractionation of guests although no such fractionation occurred. We know of at least one other research group that wasted time and resources by trying to model these apparent fractionations in vain.

The Poisson model's success tempts one to employ the generalized one in order to take mixing into account, but neither does the underlying physics support this, nor do functional fits with generalized Poisson distributions happen to fit well: average and maximum are still bound too closely. There is increasing interest in embedding large structures in noble cluster matrixes. This desires large host clusters. Correspondingly large cross sections render available vacuum technology insufficient. With the helium droplet technique's cluster sizes ever growing [25], the standard Poisson model should and can be improved analytically. Some necessary models have been derived here and have been already successfully applied to account for pre-doping by residual background gas in experiments involving large Helium droplets. Moreover, we think that our considerations will be helpful when incorporating effects like sticking coefficients and beam depletion in a rigorous way on a theoretical level.

## 10. Appendix

**Probability Density Functions (variable of interest is continuous)**

Gamma ($\Gamma$): $\quad \text{PDF}^{-1} = \langle N \rangle e^{Nd_N/\langle N \rangle} \left( Nd_N/\langle N \rangle \right)^{-d_N} \Gamma_{(d_N)}$

Exponential (EXP): $\quad \text{PDF}^{-1} = \langle N \rangle e^{N/\langle N \rangle}$

Log-Normal (LN): $\quad \text{PDF}^{-1} = N\sqrt{2\pi}\Delta_n \exp[\frac{1}{2}\left(\frac{n-\langle n \rangle}{\Delta_n}\right)^2 ]$

$\langle N \rangle = \exp\left[\langle n \rangle + \Delta_n^2/2\right]$ and $\Delta_N = \langle N \rangle \sqrt{e^{\Delta_n^2}-1}$ (or $\Delta_n^2 = \ln\left[1+1/d_N\right]$) transform between $n := \ln N$ and $N$-spaces.

Inverse Gaussian (IG): $\quad \text{PDF}^{-1}_{(d_N,\langle N \rangle)} = N\sqrt{\frac{2\pi N}{d_N \langle N \rangle}} \exp[\frac{d_N}{2N\langle N \rangle}(N-\langle N \rangle)^2]$

**Probability Distributions (variable of interest is discrete)**

Poisson: $$P_k = P_0 \bar{k}^k / k! \quad ; \quad P_0 = e^{-\bar{k}}$$

Generalized Poisson (GP): $$P_k = \frac{u(u+\lambda k)^{k-1}}{k!} e^{-u-\lambda k}$$

General expression for Poisson mixture:

$$\langle P_k \rangle = \frac{\langle \bar{k}^k \rangle}{k! G} \sum_{i=0}^{\infty} (-1)^i \frac{\langle \bar{k}^i \rangle}{i!} H \qquad (10)$$

$G = \Gamma_{(1+a)}$, $H = \Gamma_{(1+a(k+i))} / \Gamma_{(1+ai)}$ and $\langle \bar{k}^x \rangle = \Gamma_{(1+ax)} \left( \kappa / \Gamma_{(1+a)} \right)^x$ in case of the EXP;

$G = 1$, $H = \exp\left[ki\Delta^2_{\ln \bar{k}}\right]$ and $\langle \bar{k}^x \rangle = \left(\kappa \exp\left[(x-1)\Delta^2_{\ln \bar{k}}/2\right]\right)^x$ with $\Delta_{\ln \bar{k}} = a\Delta_n$

are valid for the LN being the mixing distribution.

Negative Binomial (NB):

$$\langle P_k \rangle = \frac{\Gamma_{(d_N+k)}}{k! \Gamma_{(d_N)}} \left(1 - \langle P_0 \rangle^{1/d_N}\right)^k \langle P_0 \rangle \quad ; \quad \langle P_0 \rangle = \left[d_N/(\kappa+d_N)\right]^{d_N} \qquad (11)$$

IG Poisson mixture[i] (IGP):

$$\langle P_k \rangle = \frac{e^{d_{\bar{k}}}}{k!} \sqrt{\frac{2}{\pi}} (\kappa/\Omega)^k (d_{\bar{k}}\Omega)^{(1+2k)/4} K_{\frac{1}{2}-k}\left(\sqrt{d_{\bar{k}}\Omega}\right) \qquad (12)$$

Computationally, the IGP is best gotten by recursion [26] with

$$\langle P_k \rangle = \frac{\kappa}{\Omega k}\left((2k-3)\langle P_{k-1}\rangle + \frac{\kappa d_{\bar{k}}}{k-1}\langle P_{k-2}\rangle\right).$$

---

[i] $K_c(x) = (x/2)^{-c} \int_0^\infty u^{c-1} / \left(2 \exp\left[u + x^2/4u\right]\right) du$ is the modified Bessel function of the second kind.

Manipulating the IGP needs often considering properties of the Bessel function. For instance, $\langle P_0 \rangle$ follows from the general formula via $K_{\pm 1/2}(x) = e^{-x}\sqrt{\pi/(2x)}$.

EXP-Poisson mixture for liquid drop cross section ($a = 2/3$):

$$\langle P_k \rangle = \frac{1}{k!}\left(\frac{\kappa}{\Gamma_{5/3}}\right)^k \left\{ T_{3,3,1,2} - \frac{\kappa}{\Gamma_{5/3}} T_{5,4,2,4} + \frac{1}{2}\left(\frac{\kappa}{\Gamma_{5/3}}\right)^2 T_{7,5,4,5} \right\} \qquad (13)$$

$T_{\alpha,\beta,\gamma,\delta} := {}_2F_2[\{\frac{\alpha+2k}{6},\frac{\beta+k}{3}\},\{\frac{\gamma}{3},\frac{\delta}{3}\},4\left(\frac{-\kappa}{3\Gamma_{5/3}}\right)^3]\Gamma_{\frac{\alpha+2k}{3}}$ contains the generalized

hypergeometric function ${}_2F_2[\{v,w\},\{x,y\},z]$.

Generalized Poisson written as if it is already a Poisson mixture:

$$\langle P_k \rangle_{GP} = \frac{\kappa^k}{k!}(1-\lambda)(1-\lambda+\lambda k/\kappa)^{k-1} e^{\kappa(\lambda-1)-\lambda k} \qquad (14)$$

---